\documentclass[aps,twocolumn,prb,preprintnumbers,amsmath,amssymb,superscriptaddress,floats]{revtex4-2}
\usepackage{graphicx}
\usepackage{bm}
\usepackage{float}
\usepackage{multirow}
\usepackage{array}
\usepackage{physics}
\usepackage{siunitx}
\usepackage{verbatim}
\usepackage{xcolor}
\usepackage{amsmath,lipsum}

\def\pzo{Pr$_2$Zr$_2$O$_7$}
\def\nzo{Nd$_2$Zr$_2$O$_7$}

\begin{document}

\title{Raman Scattering  Evidence for Fluctuating Kagome-Plane Moments in $\mathrm{B} \parallel [111]$ in   Pr$_2$Zr$_2$O$_7$ pyrochlore}

\author{Yuanyuan Xu}
\affiliation{Institute for Quantum Matter and Department of Physics and Astronomy, Johns Hopkins University, Baltimore, Maryland 21218, USA}

\author{Huiyuan Man}
\affiliation{Institute for Quantum Matter and Department of Physics and Astronomy, Johns Hopkins University, Baltimore, Maryland 21218, USA}
\affiliation{Institute for Solid State Physics, University of Tokyo, Kashiwa, Chiba 277-8581, Japan}

\author{Nan Tang}
\affiliation{Institute for Solid State Physics, University of Tokyo, Kashiwa, Chiba 277-8581, Japan}
\affiliation{Department of Physics, University of Tokyo, Bunkyo-ku, Tokyo 113-0033, Japan}

\author{Li Xiang}
\affiliation{National High Magnetic Field Laboratory, Tallahassee, FL 32310, USA}

\author{Sami Muhammad}
\affiliation{Institute for Quantum Matter and Department of Physics and Astronomy, Johns Hopkins University, Baltimore, Maryland 21218, USA}

\author{Komalavalli Thirunavukkuarasu}
\affiliation{Department of Physics, Florida AM University, Tallahassee, Florida 32307, USA}

\author{Dmitry Smirnov}
\affiliation{National High Magnetic Field Laboratory, Tallahassee, FL 32310, USA}

\author{Satoru Nakatsuji}
\affiliation{Institute for Solid State Physics, University of Tokyo, Kashiwa, Chiba 277-8581, Japan}
\affiliation{Department of Physics, University of Tokyo, Bunkyo-ku, Tokyo 113-0033, Japan}
\affiliation{Institute for Quantum Matter and Department of Physics and Astronomy, Johns Hopkins University, Baltimore, MD 21218, USA}
\affiliation{CREST, Japan Science and Technology Agency, Kawaguchi, Saitama 332-0012, Japan}
\affiliation{Trans-scale Quantum Science Institute, University of Tokyo, Bunkyo-ku, Tokyo 113-0033, Japan}
\affiliation{Canadian Institute for Advanced Research (CIFAR), Toronto, Ontario, Canada}

\author{Natalia Drichko}
\email[Corresponding author:]{drichko@jhu.edu}
\affiliation{Institute for Quantum Matter and Department of Physics and Astronomy, Johns Hopkins University, Baltimore, Maryland 21218, USA}

\begin{abstract}
The exotic magnetic properties of the  pyrochlore Pr$_2$Zr$_2$O$_7$ are determined by the coupling of spin, orbital, and lattice degrees of freedom. Magnetism in this material originates from the non-Kramers Pr$^{3+}$ ion, and the exotic behavior has  been  discussed both in the framework of dipole-quadrupole properties of magnetic moments and disorder splitting the non-Kramers doublet. We use magneto-Raman spectroscopy to probe the crystal electric field (CEF) excitations of  Pr$^{3+}$ in magnetic fields up to 14 T applied along the [100] and [111] crystallographic directions at 2 K. For $\mathrm{B}\parallel[100]$, the field evolution of the Raman active crystal field modes is quantitatively described by conventional Zeeman splitting of the Pr$^{3+}$ ground state doublet, including the thermally populated upper Zeeman branch at elevated temperature. For $\mathrm{}{B}\parallel[111]$, the spectra separate into responses from the triangular and kagome sublattices, and the CEF excitations of Pr$^{3+}$ on the triangular lattice  exhibit the expected Zeeman shift consistent with full moments polarization. In contrast, the CEF excitations of Pr$^{3+}$ on the  kagome lattice do not split as expected and instead broaden  strongly with increasing field. This behavior is inconsistent with a static polarized configuration and is captured phenomenologically by a motional-narrowing description in which the kagome plane moments fluctuate between Zeeman-split levels on a meV timescale. This result offers  a natural explanation for the reduced magnetization observed in $\mathrm{}{B}\parallel[111]$ and provides further evidence of the  dipole-quadrupole properties of magnetic moments. It  highlights Raman scattering as a sensitive probe of exotic spin-orbital dynamics in frustrated magnets.
\end{abstract}

\date{\today}
\maketitle

\section{Introduction}

Frustrated magnets provide a fertile source for unconventional states of matter when competing interactions  suppress conventional long-range order and stabilize highly correlated dynamical regimes.  In rare earth pyrochlores  with general formula A$_2$B$_2$O$_7$ and  rare earth atoms at site A  ~\cite{Balents2010,Broholm2020} the interplay of geometric frustration, strong spin-orbit coupling, and crystal-field anisotropy gives rise to a wide range of exotic magnetic states, including classical and quantum spin ice, emergent gauge dynamics, and multipolar ordering.~\cite{Den2000,Gardner2010,Gingras2014,Smith2024}.  Of particular importance is that low-energy degrees of freedom  are often not purely magnetic dipoles, but involve coupled dipolar and higher-rank multipolar moments~\cite{Rau2019,Smith2024}.

One of the examples of rare earth pyrochlores with exotic magnetic properties is \pzo. Pr$^{3+}$, $J$=4 is a non-Kramers ion, where a doublet ground state ensures a behavior of a system similar to S=1/2~\cite{Onoda2011,Rau2019}. Original suggestions for quantum spin ice ~\cite{Kimura2013} were complicated by the assignment of  the transverse field components  which lead to  a   split the ground state doublet in a non-Kramers ion  to disorder.~\cite{Wen2017}.   After the highest purity samples were obtained, it was confirmed that \pzo\ is a spin-orbital liquid driven by  dipole-quadrupole nature of Pr$^{3+}$ magnetic moment~\cite{Tang2020}. Recently, there appeared a theoretical suggestion, demonstrating that the quadrupole component can be dominant and lead to a spin liquid state~\cite{An2025}. In magnetic field B$\parallel$ [111] a transition between a spin-orbital liquid and fully magnetically polarized state was demonstrated at around 2~T at cryogenic temperatures~\cite{Tang2020}. In this work we have uncovered a presence of moments fluctuating in kagome plains of Pr$^{3+}$ in magnetic field B$\parallel$ [111] at 2~K. 

Rare earth pyrochlores can host an exotic magnetic state in B$\parallel$ [111] : They can be considered as two magnetic subsystems, one of them is a triangular lattice which becomes fully polarized at small magnetic fields with magnetic moments parallel to the field, allowed by the local [111] axis, and another sublattice forming kagome lattice. The latter can host kagome spin ice, where the moments are fluctuating following the ice rules~\cite{Bojesen2017}. Such dynamic kagome ice  was observed at  magnetic fields below those  inducing a fully polarized state in \nzo\, where Nd$^{3+}$ is a Kramers doublet, using neutron scattering in $<$111$>$ magnetic field~\cite{Lhotel2018}.
While not yet observed experimentally, a spin-orbital liquid on a kagome lattice of non-Kramers ions, specifically   Pr$^{3+}$, has been proposed  theoretically~\cite{Schaffer2013}. However, neutron scattering under [111] field, expected to effectively isolate  kagome planes, have not yielded any conclusive evidence~\cite{Petit2016}.

Here we demonstrate a detection of magnetic moments fluctuating in kagome planes in \pzo\ under [111] magnetic field at 2~K using Raman scattering spectroscopy.
We use magneto-Raman scattering  to directly probe the crystal electric field (CEF) excitations of Pr$^{3+}$ in Pr$_2$Zr$_2$O$_7$ under magnetic fields up to 14~T applied along $[100]$ and $[111]$. For  $\mathrm{B} \parallel [100]$, the observed spectra are well described by conventional Zeeman splitting of the crystal-field levels.
In contrast, for  $\mathrm{B} \parallel [111]$, Zeeman splitting is observed only for the moments in the triangular sublattice, fully polarized with magnetic field. The CEF excitations of the moments in kagome planes do not show Zeeman splitting, but broaden with the increase of the field values. We model this behavior as a  presence of fluctuations of magnetic moments in kagome plains between the Zeeman split levels.  Our work provides an insight in previously detected values of magnetization lower than expected ones~\cite{Petit2016,Tang2020} and suggests the importance of the quadrupolar component of the magnetic moment.

\section{Results}

\begin{figure*}[!htb]
	\includegraphics[width=\linewidth]{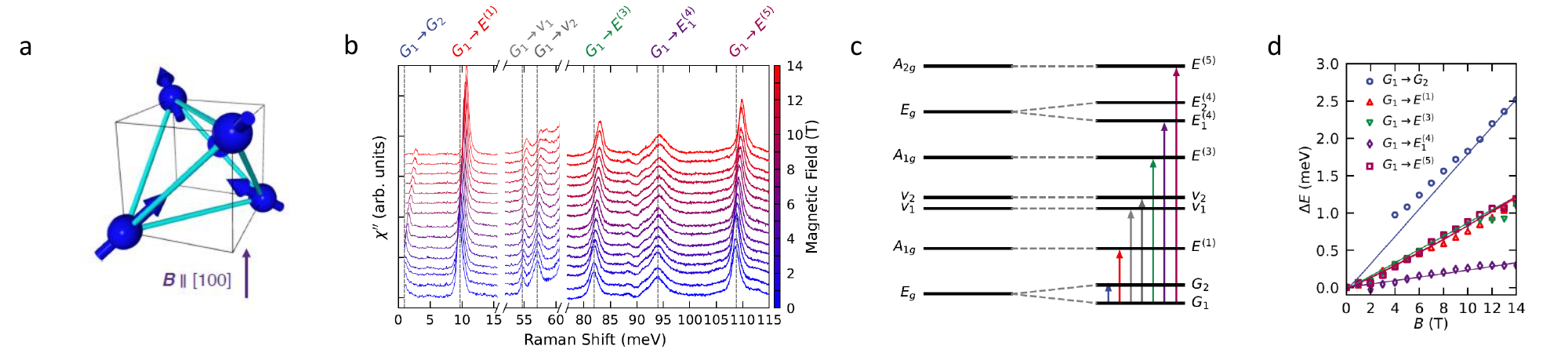 }
	\caption{(a) Schematic of Pr$^{3+}$ tetrahedron with magnetic field $\mathrm{B} \parallel [100]$. (b)  Field-dependence of unpolarized Raman spectra of Pr$^{3+}$ crystal electric field excitations  in $\mathrm{B} \parallel [100]$ at 2~K. The assignments of the excitations are marked on the top of the plot. (c) Illustration of the Zeeman effect in CEF scheme of Pr$^{3+}$, with colored arrows showing transitions marked in (b). (d) Experimental Zeeman shift of the observed crystal field excitations as a function of magnetic field E$_{CEF}$(B)-E$_{CEF}$(0~T)  (hollow markers) and respective calculations of Zeeman splitting following Eq.~\ref{Zeeman}  (solid lines).  }
    \label{Fig1}
\end{figure*}

In this work we will focus on the Raman scattering of the crystal field excitations of  Pr$^{3+}$. Pr$_2$Zr$_2$O$_7$ has a cubic pyrochlore structure of space group $Fd\bar{3}m$, where the Pr$^{3+}$ ion is located at the site of $D_{3d}$ symmetry. The ground state $^3H_4$ of the $4f$ electrons of Pr$^{3+}$ ($J$=4) is split into $2E_g + 2A_{1g} + A_{2g}$ multiplets. One of the higher energy $E_g$ CEF levels at around 55~meV is split into a pair of vibronic states due to the phonon-CEF coupling \cite{Xu2021}. The non-Kramers $E_g$ ground state doublet splitting, possibly originating from the displacement of the Pr$^{3+}$ ions, is suppressed at 2~K \cite{Wen2017, Martin2017, Xu2021}. In the absence of the magnetic field, six CEF transitions from the CEF ground state to the excited CEF states and vibronic states are assigned based on the previous work \cite{Xu2021}.

In magnetic field, we expect Zeeman splitting of the CEF levels, determined by the following equation

\begin{equation}
	\begin{split}
	\mathcal{H}_{\mathrm{Zeeman}} &= -\mu_\mathrm{B}g_J\vb{B}\cdot\vb{J}\\
	 	&= -\mu_\mathrm{B}g_J(BJ_x\sin\theta + BJ_z\cos\theta),
	\end{split}	
 \label{Zeeman}
\end{equation}
where $g_J = 4/5$ is the Land\'e g-factor. Here $J_z$ will provide the main contribution to the Zeeman splitting in this Ising system, while  $J_x$ is a transverse component expected in \pzo \cite{Wen2017,Tang2020}.

The CEF Hamiltonian for Pr$^{3+}$ in $D_{3d}$ local environment in the LS coupling scheme takes the form of \cite{Hutchings1964},
\begin{equation}
	\begin{split}
		\mathcal{H}_{\mathrm{CEF}} = &B_2^0 O_2^0 + B_4^0O_4^0 + B_4^3 O_4^3 +\\ &B_6^0 O_6^0 + B_6^3 O_6^3 + B_6^6 O_6^6,
	\end{split}	
\end{equation}
where $O_n^m$ are the Stevens operators \cite{Stevens1952}, $B_n^m$ are the CEF parameters given in Table.~\ref{table:CEF_parameters}.  Keeping this basic information in mind, we can discuss  the experimental data acquired using  B$\parallel$ [100] and B$\parallel$ [111].

\begin{table}[H]
	\caption{CEF parameters $B_m^n$ (in $\mu$eV) for Pr$_2$Zr$_2$O$_7$ determined by the previous work \cite{Xu2021}.}
	\label{table:CEF_parameters}
	\centering
	\begin{ruledtabular}
	\begin{tabular}{
		>{\centering\arraybackslash}l
    	>{\centering\arraybackslash}p{0.12\linewidth}
		>{\centering\arraybackslash}p{0.12\linewidth}
		>{\centering\arraybackslash}p{0.12\linewidth}
		>{\centering\arraybackslash}p{0.12\linewidth}
		>{\centering\arraybackslash}p{0.12\linewidth}}
		$B_2^0$ & $B_4^0$ & $B_4^3$ & $B_6^0$ & $B_6^3$ & $B_6^6$\\
		\hline
		 -650.2 & -33.08 & -460.5 & 0.268 & 1.891 & -1.987
	\end{tabular}
	\end{ruledtabular}
\end{table}

\subsection{Spectra of crystal field levels of \pzo\ in  $\mathrm{B} \parallel [100]$}

\begin{figure*}[!htb]
	\includegraphics[width=\linewidth]{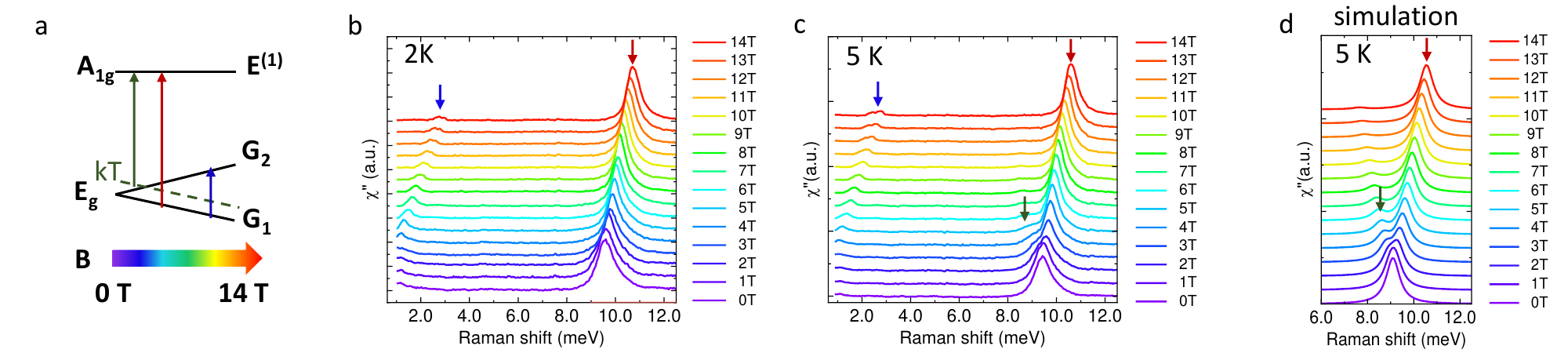 }
	\caption{Raman active transitions from the ground state to the first excited state of  Pr$^{3+}$ in \pzo\ in B$\parallel$ [100]. (a) Scheme of the Zeeman splitting of the ground state of  Pr$^{3+}$ in \pzo\ in B$\parallel$ [100], illustrating the importance of thermal population of the levels, possible Raman active transitions marked with arrows. (b) Experimental Raman scattering spectra of the transitions from the ground state to the first excited state at T=2~K in B$\parallel$ [100] for magnetic fields from 0 to 14 T. Transition $G_1 \rightarrow E^{(1)}$ marked with red arrow, $G_1 \rightarrow G_2$ observed at B $>$ kT is marked with blue arrow.  (c) Experimental Raman scattering spectra of the transitions from the ground state to the first excited state at T=5~K in B$\parallel$ [100] for magnetic fields from 0 to 14 T.  Transition $G_1 \rightarrow E^{(1)}$ marked with red arrow, $G_2 \rightarrow E^{(1)}$ observed at B $<$ kT is marked with green arrow,  $G_1 \rightarrow G_2$ observed at B $>$ kT is marked with blue arrow. (d) Simulated spectra for T=5~K  in B$\parallel$ [100]  for magnetic fields from 0 to 14 T.  Transition $G_1 \rightarrow E^{(1)}$ marked with red arrow, $G_2 \rightarrow E^{(1)}$ observed at B $<$ kT is marked with green arrow.}
    \label{Fig001GS}
\end{figure*}

 Fig.~\ref{Fig1} shows the magnetic field-dependence of the Raman scattering spectra of CEF excitations in Pr$_2$Zr$_2$O$_7$ at 2~K in the energy range of 1~to 115~meV with magnetic field B$\parallel$[100] up to H= 14~T. The phonons, which are known to involve oxygen movement~\cite{Xu2022phonons} are not shown for clarity.
 The basic tendency we observe is a linear blue shift of the excitations on the increase of the  magnetic field (Fig.~\ref{Fig1}b), which can be also followed in Fig~\ref{Fig1}(d), where the shift dependent on magnetic field $\Delta E(B) = E(B) - E(0)$  vs B at 2 K is plotted  for these excitations (marked with hollow markers).

In order to interpret the spectra and calculate Zeeman splitting, we consider the fully polarized state at B $\parallel$ [100]. In this configuration, all four Pr$^{3+}$ ions found on the corners of tetrahedra in the \pzo\ structure have the same angle  $\theta = 54.7^\circ$ between the $\langle 111 \rangle$-local easy axis and the B-field direction [100]  (Fig.~\ref{Fig1}(a)). Under the Zeeman effect, the CEF ground state doublet is split into the lower-energy state $G_1$ and the higher-energy state $G_2$.   The scheme of CEF Zeeman splitting and transitions from $G_1$ to high energy states $G_2, E_1^{(1)}, v_1, v_2, E^{(3)}, E_1^{(4)}$, and $E^{(4)}$ are  indicated by the dashed gray lines in Fig.~\ref{Fig1}(b) and the colored arrows in Fig.~\ref{Fig1}(c). Table.~\ref{table:100_transitions} presents the energies and linewidths of these transitions under 14 T.

Calculated  Zeeman splitting describes well the shift of the CEF frequencies with field from about 2~T up to 8~T ( Fig.~\ref{Fig1}(c), solid lines for calculated values). The best  agreement with the calculations is obtained when we added a transverse field term $\Delta_i = 0.3$~meV , suggested in Ref. \cite{Wen2017} as a result of the effect of disorder on magnetic moments $\mathcal{H}_\mathrm{t} = \sum_i \Delta_i \sigma_i^x$ .
 We find the presence of the transverse field has little effect on the frequencies of $G_1 \rightarrow G_2$ and $G_1 \rightarrow E_1^{(4)}$ transitions but improves the consistency between the experimental data and calculated values for $G_1 \rightarrow E^{(1)}$, $G_1 \rightarrow E^{(3)}$ and $G_1 \rightarrow E_1^{(5)}$ transitions.   The CEF at 94 meV, which is expected to split into the lower-energy level $E_1^{(4)}$ and the upper-energy level $E_2^{(4)}$ is exceptionally broad  (about 3.7 meV) compared to the linewidth of other excitations (about 1 meV), which apparently obscures the effect of Zeeman splitting.

We next focus our attention on the details of the evolution of the ground state doublet under applied magnetic field. Similar to the zero-field case~\cite{Xu2021}, the line shape of the transitions from the ground state doublet to the first excited state $G_1$, $G_2 \rightarrow E^{(1)}$ found at around 9 meV is determined as I($\omega$,T)=$B_{n,k} \int \rho(\omega',H)L(\omega_i-\omega')d\omega'$.  Here  $\rho(\omega',H)$ is the density of states of the ground state doublet dependent on H-field according to the Zeeman splitting (Eq.~\ref{Zeeman}), and $L(\omega_i-\omega')$ is the Lorenzian line shape of $E^{(1)}$ singlet first excited state, which stays constant with magnetic field. This allows us an insight into the line shape of the transitions at low magnetic fields, where the system is not fully polarized.

Upper level of the ground state doublet split by Zeeman splitting is populated thermally at lower  magnetic fields.  The thermal population is described as  $N$(B,T) = $e^{-\frac{\hbar2\Delta\omega_Z}{k_BT}}/(1+e^{-\frac{\hbar2\Delta\omega_Z}{k_BT}})$, where Zeeman splitting $\Delta\omega_Z$ is defined by Eq.~\ref{Zeeman}. 
The effect  of thermal population of  $G_2$ is easily followed in the experimental spectra  at 5~K (Fig.~\ref{Fig001GS} c) where both transitions are observed  between 2 and 7 ~T:  $G_2 \rightarrow E^{(1)}$ shifting to lower frequencies, and  $G_1 \rightarrow E^{(1)}$ shifting to higher frequencies at the function of H-field in agreement with the expectations (Fig.~\ref{Fig001GS}a). Here the lower field at which  $G_2 \rightarrow E^{(1)}$ is observed  is determined by the possibility to resolve these two transitions limited by their natural width,  and the higher field depends on temperature, as it is determined by the thermal population. The simulation, which takes into account the Zeeman splitting (Eq.~\ref{Zeeman}), thermal population $N$(B,T)  for 5~K reproduces the spectra well (compare Fig.~\ref{Fig001GS} d and c). For the simulations, we use the linewidth of 0.7~meV, according to the FWHM of the excitation at B=0~T, and ignore the fact that the line width decreases down to 0.4~meV at high fields. This decrease of about 0.3~meV is in agreement with the broadening which would be introduced by the disorder in a sample, estimated in Ref.~\cite{Wen2017}. The thermal population of G$_2$ will be also affecting the lineshape of the transitions to the higher CEF levels, however the splitting between G$_1$ and G$_2$ at which the population is still detectable is small compared to the energies and width of the transitions to the higher levels, and will be only detectable as narrowing of the lines at H-fields above which G$_2$ is depopulated.

\begin{table}[H]
	\caption{Energies and linewidths of the transitions from the lower-energy ground state $G_1$ to various higher energy states under 14~T [100]~magnetic field.}
	\label{table:100_transitions}
	\centering
	\begin{ruledtabular}
	\begin{tabular}{
		>{\centering\arraybackslash}l
    	>{\centering\arraybackslash}p{0.3\linewidth}
    	>{\centering\arraybackslash}p{0.3\linewidth}}
		Transition & Energy (meV) & Linewidth (meV)\\
		\hline
		 $G_1 \rightarrow G_2$ & 2.5 & 0.5 \\
		 $G_1 \rightarrow E_1^{(1)}$ & 10.5 & 0.5 (4 cm-1)\\
		 $G_1 \rightarrow v_1$ & 55.2 & 0.7\\
		 $G_1 \rightarrow v_2$ & 57.4 & 1.1\\
		 $G_1 \rightarrow E^{(3)}$ & 82.9 & 1.5\\
		 $G_1 \rightarrow E_1^{(4)}$ & 94.2 & 3.7\\
		 $G_1 \rightarrow E^{(4)}$ & 109.9 & 1.3
	\end{tabular}
	\end{ruledtabular}
\end{table}


Above 11 T the band of $G_1 \rightarrow G_2$ transition broadens into a double excitation. At these magnetic field the field dependence of Zeeman splitting for higher-energy transitions start to deviate from the linear dependence, see Fig.~\ref{Fig001GS}d. The deviations from Zeeman splitting at high fields are small, below 10~\% of the splitting itself,  and can be only observed with the high energy resolution available by the Raman scattering spectroscopy.
The detailed  discussion of these small deviations from Zeeman splitting is out of scope of this work.   It is possible, that high magnetic fields allow to resolve slightly differently oriented domains, however, their orientation would be within 6 degrees from each other, as estimated from the observed slitting. Another possibility is a deformation of the crystal environment of Pr$^{3+}$ in magnetic field above 11~T. Such changes would be too weak to be detected through the change in phonon excitations,  but would be observed in CEF excitations, due to their high sensitivity to the local symmetry.



\subsection{Spectra of the ground state crystal field levels of \pzo\ in  $B \parallel [111] $}

\begin{figure*}[!htb]
	\includegraphics[width=\linewidth]{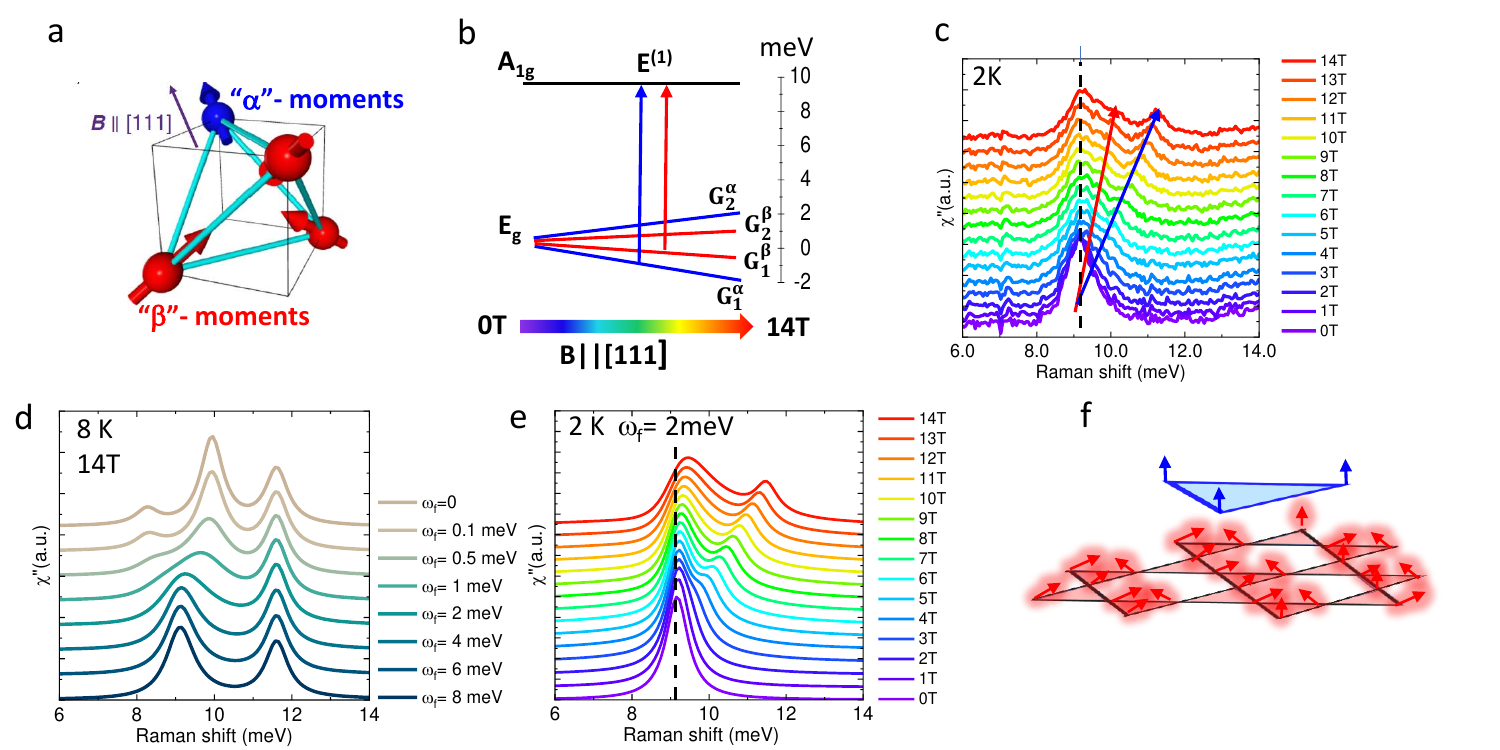}
	\caption{ (a) The scheme of Pr$^{3+}$ tetrahedron under magnetic field $\mathrm{B} \parallel [111]$. Two types of the Pr$^{3+}$ magnetic moments in respect to the magnetic field direction are marked: $\alpha$ (blue arrows) and $\beta$ (red arrows). (b) Sketch of the Zeeman effect of the transitions from the CEF ground states to the first excited state for Pr$^{3+}_\alpha$ (blue) and Pr$^{3+}_\beta$ (red) ions. (c) Experimental Raman scattering spectra of the transitions from the ground state to the first excited state at T=2~K in B$\parallel$ [111] for magnetic fields from 0 to 14 T.  (d) Dependence of $G_1^\alpha, G_1^\beta \rightarrow E^{(1)}$ transitions at 8~K and 14~T on the  fluctuation frequency $\omega_f$ according to the simulation using the motional narrowing model Eq~\ref{Kubo}. (e) Simulated spectra of $G_1^\alpha, G_1^\beta \rightarrow E^{(1)}$ transitions at 2~K under magnetic field $\mathrm{B} \parallel [111]$ for magnetic fields from 0 to 14 T using motional narrowing model (Eq.~\ref{Kubo}) with static case for $\alpha$ magnetic moments polarized to the filed and  fluctuation frequency $\omega_f$= 2~meV for $\beta$-moments. (f) Magnetic structure of a pyrochlore in $\mathrm{B} \parallel [111]$, where $\alpha$ spins form trigonal planes, with local [111] axis is parallel to magnetic field, and $\beta$- moments form kagome planes with local [111] axis at an angle to the field and are found to be fluctuating at 2~K.   }
	\label{Fig111}
\end{figure*}

 In the  $\mathrm{B} \parallel [111]$ geometry, as illustrated in the scheme in Fig.~\ref{Fig111}a, there are two distinct magnetic sites of Pr$^{3+}$ in a tetrahedron, distinguished by the orientation of the local Ising easy axis relative to the $H$-field direction $[111]$.
In a fully polarized state, magnetic moments marked in blue (Pr$^{3+}_\alpha$) are parallel to the $[111]$ field direction, while the moments marked in red (Pr$^{3+}_\beta$) form an angle of $70.5^\circ$ with respect to the field axis (see Fig.~\ref{Fig111}a). Thus, in a fully polarized state, the \pzo\ magnetic structure, similar to other pyrochlores~\cite{Ross2009,Lhotel2018}, would present planes alternating along the [111] direction: triangular lattice planes of Pr$^{3+}_\alpha$ moments fully parallel to the $H$-field and kagome planes of Pr$^{3+}_\beta$ moments.

In our discussion of Raman scattering in   $\mathrm{B} \parallel [111]$, we will focus on the transition from the ground state to the first excited state $G_1, G_2 \rightarrow$ $E^{(1)}$ (Fig.~\ref{Fig111}b). Fig.~\ref{Fig111}c presents the Raman spectra of \pzo\ in the range from 6 to 14 meV at 2 K in magnetic fields up to 14 T applied along the $[111]$ direction.
The calculations of the Zeeman splitting of the ground state level following Eq.~\ref{Zeeman} for the fully polarized state as a function of $B$-field provide the frequencies of two sets of excitations $G_1^{\alpha}, G_1^{\beta} \rightarrow$ $E^{(1)}$ as a function of field, with the difference in frequency corresponding to the angle between the $B$-field and the local Ising axis (see Eq.~\ref{Zeeman}). We mark the expected positions of the excitations vs $B$-field on top of the observed spectra in Fig.~\ref{Fig111}c, with the blue arrow marking the expected energies of the $G_1^{\alpha} \rightarrow$ $E^{(1)}$ transition, and the red arrow showing the expected $G_1^{\beta} \rightarrow$ $E^{(1)}$ energies.

The excitation corresponding to the transition $G_1^{\alpha} \rightarrow E^{(1)}$ from the ground state of the Zeeman-split level of Pr$^{3+}_\alpha$ can be easily identified in the experimental spectra. It has a well-defined line shape, and its frequency dependence on $B$-field is in agreement with the calculated Zeeman splitting.
Unexpectedly, the rest of the spectral weight is contained in an excitation that does not show a shift with magnetic field, but only an increase in the  linewidth of the asymmetric line.

The agreement of the excitation frequency of the Pr$^{3+}_\alpha$ CEF with Zeeman splitting calculations confirm that our data are robust in terms of the size of magnetic field, measured Zeeman splitting, and orientation of the sample. Assuming the same probability of the transitions  $G_1^{\alpha}, G_1^{\beta} \rightarrow$ $E^{(1)}$  for Pr$^{3+}_\alpha$ and Pr$^{3+}_\beta$, we would expect S(alpha)/S(beta)= 1:3= 0.33, where S is the spectral weight of the respective Raman line S=$\int \chi''(\omega)\delta\omega$. It is somewhat challenging to separate two excitations, since this involves assumptions about the shape of the asymmetric excitation. The lowest limit for the ratio for the data for  14 T and 2 K   presented in Fig.~\ref{Fig111} yields  S($\alpha$)/(S(total)-S($\alpha$))=0.247.  (the SW calculation is shown in the supporting information) This estimation suggests that at least 70\% of the signal from the  sample is produced by the exactly oriented domains without accounting for structural disorder~\cite{Wen2017}. Such disorder cannot explain the substantial deviation of the experimental response of Pr$^{3+}_\beta$ spins from Zeeman splitting.


In cubic pyrochlores with $\mathrm{B} \parallel [111]$, a crucial difference emerges between the behavior of the triangular lattice of Pr$^{3+}_\alpha$ ions, whose local Ising axes are parallel to $[111]$ and thus become fully polarized at small fields, and the moments residing in the kagome planes (Pr$^{3+}_\beta$). At magnetic field strengths below those required to induce full polarization, a dynamic kagome ice regime for the $\beta$ moments has been observed, for example, in Nd$_2$Zr$_2$O$_7$~\cite{Lhotel2018}. In this regime, moments within the kagome planes fluctuate between configurations aligned with $\mathrm{B} \parallel [111]$ (``up'' states) and those aligned in the opposite direction (``down'' states), while obeying the ice rules. Such fluctuations would affect the line shape of excitations between Zeeman-split bands measured by Raman scattering, provided the fluctuation frequency is sufficiently high to substantially reduce the excited-state lifetimes. In this case, the associated linewidth would increase beyond that determined by disorder or the natural lifetime.

\subsection{Description of E$_g \rightarrow$ A$_{1g}$ transition in magnetic field in terms of motional narrowing}

The simplest model to describe the line shape determined by such fluctuation is so-called the ``motional narrowing''~\cite{Kubo1969stochastic}.  This model  describes a line shape which arises due to stochastic fluctuations between two states with a fluctuation rate $\omega_f$:
\begin{widetext}
\begin{equation}
I(\omega) \propto Re \left[ \begin{pmatrix}a_{up} \\ a_{down} \end{pmatrix} \begin{bmatrix}
i(\omega-(\omega_m-\Delta\omega_Z)+\omega_f /2+\Gamma/2) & -\omega_f/2\\
 -\omega_f/2 & i(\omega-(\omega_m+\Delta\omega_Z)+\omega_f /2+\Gamma/2)
\end{bmatrix} ^{-1}	(a_{up}, a_{down})) \right]
\label{Kubo}
\end{equation}
\end{widetext}
 In the case of Pr$^{3+}$ magnetic moments in kagome planes of \pzo, we assume that  two states are moments in the ``up'' ground state, corresponding to the fully polarized configuration, and the excited ``down'' state.  The frequencies of excitations from the ``up'' ground state to the first excited state G$_1^{\beta} \rightarrow$ E$^{(1)}$ ($\omega_\mathrm{up}$),   and ``down'' states  G$_2^{\beta} \rightarrow$ E$^{(1)}$  ($\omega_\mathrm{down}$) are determined by Zeeman splitting  $\frac{\omega_\mathrm{down}-\omega_\mathrm{up}}{2}=\Delta\omega_\mathrm{Z} (B)$ (see Eq~\ref{Zeeman}). $\omega_m=\frac{\omega_\mathrm{up}-\omega_\mathrm{down}}{2}$ denotes the mean frequency of the two excitations, in our case $\omega_m$=$\omega_{B=0}$, the transition frequency in the absence of a magnetic field. The parameter $\Gamma$ represents the natural linewidth, determined by the lifetime of the excited level in the static limit. The quantities $a_\mathrm{up}$ and $a_\mathrm{down}$ correspond to the intensities of the transitions G$_1^{\beta} \rightarrow$ E$^{(1)}$ and G$_2^{\beta} \rightarrow$ E$^{(1)}$, respectively.

To illustrate how the line shape of the excitations G$_1^{\beta}$, G$_2^{\beta} \rightarrow$ E$^{(1)}$ depends on the fluctuation frequency $\omega_f$ between these two states, we present a calculation for the $G_{1,2}^{\alpha}$, $G_{1,2}^{\beta} \rightarrow$ $E^{(1)}$ transitions in \pzo\ at $B = 14$~T and $T = 8$~K, see Fig.~\ref{Fig111}d. The relative intensities $a_\mathrm{down}$ and $a_\mathrm{up}$ are determined by the thermal populations of the Zeeman-split levels.
 For $\omega_f$ = 0 (static case) the line shapes and energies are determined solely by Zeeman splitting and natural widths of the excitations.
As the fluctuation frequency $\omega_f$ for Pr$^{3+}_\beta$ moments increases, the excitation lines broaden due to a reduction in the lifetime of the levels. When $\omega_f$ becomes comparable to $\Delta\omega_\mathrm{Z}$, the two peaks merge into a single peak, with a width determined by the exchange frequency $\omega_f$ and a line shape determined by the transition frequencies and intensities, which in this example depend on the thermal populations of the  levels (Fig.~\ref{Fig111}b). The temperature $T = 8$~K was chosen as an example to ensure appreciable thermal population of the G$_2^{\beta}$ level, N$_{G2}$(B) = $e^{-\frac{\hbar2\Delta\omega_Z(B)}{k_BT}}/(1+e^{-\frac{\hbar2\Delta\omega_Z(B)}{k_BT}})$.
This simplified model neglects the possible collective nature of kagome ice excitations; however, it remains valid for crystal-field excitations, which are inherently local in character. This approach has been previously demonstrated to successfully reproduce Raman scattering line shapes when applied to spectroscopic studies of fluctuations, such as charge fluctuations in organic Mott insulators~\cite{Yakushi2012_Crystals,Hassan2018}.

As the next step, we apply this model to reproduce the low-frequency spectra of \pzo\ at $T = 2$~K in $\mathrm{B} \parallel [111]$ (Fig.~\ref{Fig111}c) as a function of magnetic field. We use Eq.~\ref{Kubo} with the intensities $a_\mathrm{up}$ and $a_\mathrm{down}$ determined by the thermal populations of the levels and the Zeeman splitting calculated according to Eq.~\ref{Zeeman}. The fluctuation frequency $\omega_f$ is the only adjustable parameter used to describe the broad line of the G$_{1,2}^{\beta} \rightarrow$ E$^{(1)}$ transition (see Fig.~\ref{Fig111}e), while the G$_{1,2}^{\alpha} \rightarrow$ E$^{(1)}$ transition is modeled with $\omega_f = 0$. The broad, asymmetric line shape of the  G$_{1,2}^{\beta} \rightarrow$ E$^{(1)}$ excitation is well reproduced with $\omega_f = 2$~meV (see Fig.~\ref{Fig111}e). However, in the modeled spectra, the maximum intensity of the excitation shifts to higher frequencies by approximately 0.3~meV as the magnetic field increases to 14~T, in contrast to the experimental spectra, which exhibit minimal shift (Fig.~\ref{Fig111}c). Better agreement regarding the absence of frequency shift could be achieved through small variations of $\omega_f$, introduction of a broader distribution of fluctuation frequencies, or field-dependent modification of the natural linewidth. However, given the simplicity of the model, it is hard to justify such adjustments.   This simple model illustrates the main result: the moments in the kagome planes behave qualitatively differently not only from the moments in the triangular planes under $\mathrm{B} \parallel [111]$, but also from the moments under $\mathrm{B} \parallel [100]$. The key difference lies in the reduced lifetime, which is shortened by fluctuations with frequencies on the order of the Zeeman splitting between the levels.

\section{Discussion}

Theoretical predictions~\cite{Moessner2003,Isakov2004,Bojesen2017} and experimental studies of Dy$_2$Ti$_2$O$_7$~\cite{Sakakibara2003,Tabata2006}, Ho$_2$Ti$_2$O$_7$~\cite{Fennell2007}, and Nd$_2$Zr$_2$O$_7$~\cite{Lhotel2018} demonstrate that the kagome ice state emerges when the degeneracy of the pyrochlore lattice is lifted by a magnetic field applied along $\mathrm{B} \parallel [111]$. Under this field orientation, the pyrochlore lattice splits into a triangular sublattice of magnetic moments, which are polarized along the $[111]$ direction by an infinitely small magnetic field, and kagome planes, which can host kagome spin ice at intermediate fields. In particular, a dynamic kagome spin ice state was demonstrated at intermediate fields in Nd$_2$Zr$_2$O$_7$ due to the presence of dipole-octupole moments~\cite{Lhotel2018}. The $\mathrm{B} \parallel [111]$ magnetic-field-induced kagome spin ice in pyrochlores is observed at cryogenic temperatures $T < 1$~K and is typically evidenced by the observation of a magnetization plateau and characteristic neutron scattering signatures at the relevant magnetic fields. At these cryogenic temperatures, $\mathrm{B}_2 > 2J_z$~\cite{Carrasquilla2015,Bojesen2017} corresponds to the field of the spin-flip transition, above which the system becomes fully polarized.



The study of the phase diagram of \pzo\ by magnetostriction as a function of magnetic field along $[111]$ reveals a sharp transition from a 3D spin-orbital liquid directly to a fully spin polarized state at 2~T below approximately 0.06~K~\cite{Tang2020}. At higher temperatures, which are still below the regime where our Raman measurements are performed, this work suggests a crossover between these two states, with the crossover magnetic field increasing with temperature. The quantum spin-orbital liquid behavior is explained by quadrupolar coupling. It is interesting to compare the temperature regime of our measurements (2~K) with the expectations for the higher temperature regime of the phase diagram in Ref.~\cite{Tang2020}: a rough linear extrapolation of the crossover line to higher temperatures would place the crossover at magnetic fields above 10~T, suggesting that magnetic moment fluctuations persist at magnetic fields comparable to the highest fields in our measurements.

Previous measurements of \pzo\ at 0.09~K and above~\cite{Petit2016} revealed that the magnetic moment in a [111] magnetic field up to 8~T is lower than expected for the fully polarized state, in agreement with Ref.~\cite{Tang2020} and our own magnetization data.
Neutron diffraction data obtained in  $\mathrm{B} \parallel [111]$ could not be uniquely refined, and the authors consequently suggested that magnetic moment for both triangular and kagome planes increases with field  while preserving different amplitudes~\cite{Petit2016}.
According to our Raman results, which can distinguish the CEF excitations and Zeeman splittings  of triangular and kagome planes, the moments in triangular lattice layers are fully polarized in field  at least above 2~T at 2~K, where we can distinguish separately G$^{\alpha}_1 \rightarrow$ E$^{(1)}$ transition, and confirm that its frequency is in agreement with Zeeman splitting of the fully polarized moments.

We therefore suggest that the fluctuating moments in kagome planes are responsible for the lower values of magnetization.
While Raman scattering on CEF excitations cannot distinguish between local and collective behavior, the spectroscopic evidence of the fluctuations of Pr$^{3+}$ magnetic moments in kagome planes up to the highest measured field of 14~T is robust. The frequency is much higher than the characteristic time scale of neutron diffraction, thus it is natural that neutron diffraction  did not produce a definite result~\cite{Petit2016}. The observed Raman scattering of this CEF excitation is more complex  than a modeled spectrum, which includes single frequency fluctuations between  the Zeeman-split levels  with respective thermal population. Interestingly, at 14~T the `upper'' $G^{\beta}_2$ level has negligible thermal population, since Zeeman splitting at 14 T for $\beta$ moments is about 2$\Delta \omega_Z$= 1.4 meV, and would not affect the spectra in a static case.  We cannot exclude thermal effects completely, since at cryogenic temperatures the system becomes fully polarized at fields above 2~T~\cite{Tang2020}. Importantly, the observed high frequency fluctuations are present  only for the moments in kagome planes in $\mathrm{B} \parallel [111]$.  In the configuration with   $\mathrm{B} \parallel [100]$, transitions from the thermally populated levels, as well as those between the levels in a Zeeman-split doublet, are understood well in a static $\omega_f$=0 regime.

It is worth noting, that \pzo\ has been suggested as one of the candidate pyrochlores to observe dynamic kagome ice in a  [111]  magnetic field, with the interactions responsible for 3D quantum spin ice leading to the dynamic kagome ice~\cite{Carrasquilla2015}. Quadrupolar interactions~\cite{Onoda2010}, which are a proposed origin of the spin orbital liquid observed at cryogenic temperatures~\cite{Tang2020}, are a natural candidate for the interactions leading to the fluctuations of magnetic moments in the kagome planes at 2~K observed in this work. While such state is expected and has been observed at intermediate magnetic fields in other materials, magnetic  moments in kagome plains fluctuating at 14 ~T and with the high frequency of 2 meV is an unconventional result, taking into account that most of the energy scale of the interactions are expected to be on the order of  1 K   \cite{Wen2017,Tang2020}, though exact values are mostly unknown~\cite{An2025}.  The fluctuations which extend to such high magnetic fields suggest high values of the quadrupolar interactions.  There was recently a theoretical proposal that quadrupolar component can exceed exchange scattering~\cite{An2025}, and result in quantum spin liquid states beyond quantum spin ice, assumed previously for \pzo.  It would be interesting to consider how our result is related to this new framework.
Finally, we would like to comment on the suggested Raman detection of
spin-orbital liquids on a kagome lattice for non-Kramers atoms, where
the broadening of an E$_g$ phonon is expected as a
consequence of such state~\cite{Schaffer2013}. Such broadening is indeed observed in
\pzo\ as compared, for example, to Nd$_2$Zr$_2$O$_7$~\cite{Xu2022phonons,Muhammad2025},
but it extends over a broad temperature range beyond the regime where
spin-orbital fluctuations occur.


To summarize, our Raman scattering study of the crystal electric field excitations of \pzo\ at 2~K and in magnetic field up to 14~T applied in $\mathrm{B} \parallel [100]$ and $\mathrm{B} \parallel [111]$ directions demonstrate, that while crystal electric field excitations are well-described by  Zeeman splitting in  $\mathrm{B} \parallel [100]$, in  $\mathrm{B} \parallel [111]$ Raman spectra reveal  the presence of the fluctuations of magnetic moments in the kagome planes, when the degeneracy of the pyrochlore lattice is relieved by the application of  $\mathrm{B} \parallel [111]$ field.  We relate this behavior to the large quadrupolar interactions in this material. This result provides the insight to the origin of a discrepancy in magnetization values of \pzo\ in [111] field.  Both the results of this work and previously published results on \pzo, suggested  that in $\mathrm{B} \parallel [111]$ field magnetization of \pzo\ up to at least 8~T does not show magnetic moment expected from the fully polarized state.

Such possibility to separate contributions of triangular and kagome planes and probe dynamics on meV-related time scale   demonstrates the power of Raman spectroscopy to probe exotic emergent states of matter.




\section{Methods}

Pr$_2$Zr$_2$O$_7$ single crystals were grown using the floating zone method~\cite{Koohpayeh2014}. The crystals were aligned using X-ray diffraction and cut to enable Raman scattering measurements from surfaces parallel to the $[100]$ and $[111]$ crystallographic planes. Magneto-Raman experiments were performed at the National High Magnetic Field Laboratory (NHMFL) using a Quantum Design PPMS cryostat equipped with a 14~T magnet and a home-built insert for Raman scattering measurements. Raman scattering was excited using an unpolarized 532~nm laser with the excitation power maintained at approximately 400~$\mu$W to minimize laser heating. The scattered light was collected in backscattering geometry with the magnetic field applied in Faraday configuration and directed to a Princeton Instruments monochromator equipped with a liquid-nitrogen-cooled CCD detector.

\begin{figure}[!htb]
	\includegraphics[width=\linewidth]{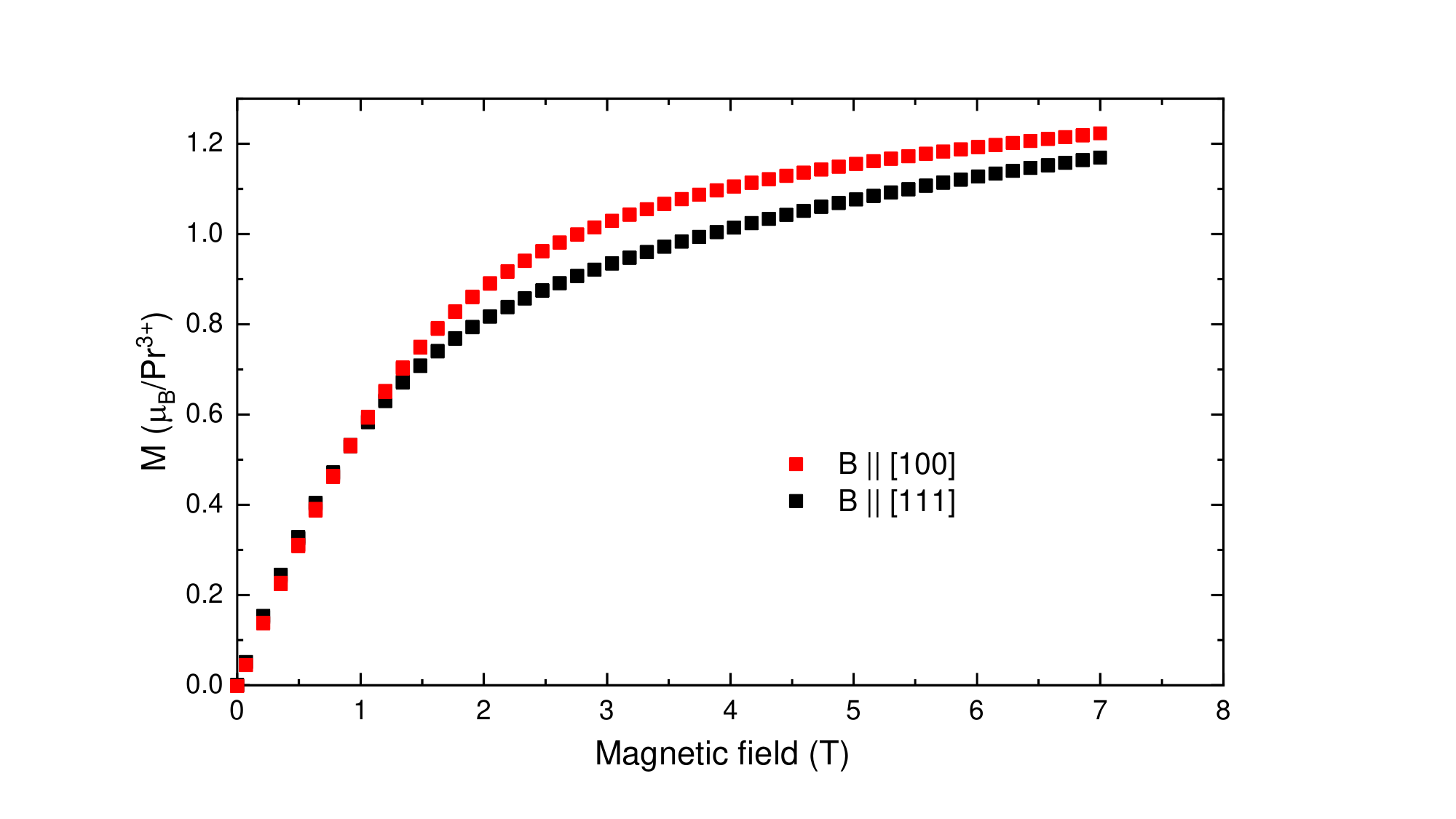}
	\caption{ Magnetization of \pzo\ in $\mathbf{B} \parallel [100]$ and $\mathbf{B} \parallel [111]$  at 2~K.}
	\label{FigM}
\end{figure}

Samples used for Raman scattering measurements were characterized by magnetization measurements at 2~K in magnetic field up to 7~T (see Fig.~\ref{FigM})  performed using a Quantum Design MPMS.

\newpage



\newpage

\section{Acknowledgment}
The authors thank O.~Tchernyshyov, C. Broholm, J.~Zhang, Y.~Luo, H. Chen for fruitful discussions.

\section{Funding}
Work at JHU  was supported as part of the Institute for Quantum Matter, an Energy Frontier Research Center funded by the U.S. Department of Energy, Office of Science, Basic Energy Sciences under Award No. DE-SC0019331 and  IQM Bridge funding from the William H. Miller III department of Physics and Astronomy. We acknowledge the support of JST-ASPIRE (JPMJAP2317, JPMJAP2512)  and "JSPS-KAKENHI (JP25H01250). D.S. acknowledges the support from the U.S. Department of Energy (No. DE-FG02-07ER46451) for magneto-Raman spectroscopy measurements performed at the National High Magnetic Field Laboratory, which is supported by National Science Foundation through DMR-1644779 and DMR-2128556, and the State of Florida.

\section{Authors Contributions}
YX performed measurements and data analysis, participated in writing the manuscript; , H.M.and N.T. grew and characterized samples, S.M. performed part of data analysis; L.X.,  D.S. facilitated and supervised the measurements; K.T. supervised he measurements, participated in writing the manuscript; S.N. supervised growth and characterization of the samples, participated in writing the manuscript; N.D. conceived the project, performed data analysis, wrote the manuscript.
\\
\\
Ethics declaration: not applicable.
\\
\\
The Authors declare no Competing Interests.
\\
\\
Data are available from the corresponding author on a reasonable request.

\bibliography{PZO}

\end{document}